\documentstyle[prb,aps,epsfig,psfig]{revtex}
\begin{document}

\sloppy

\title{Discrete breathers in classical spin lattices}

\author{Y. Zolotaryuk and S. Flach}
\address{Max-Planck-Institut f\"ur Physik komplexer Systeme,
N\"othnitzer Str. 38, D-01187 Dresden, Germany}
\author{V. Fleurov}
\address{School of Physics and Astronomy, Raymond and Beverly Sackler
Faculty of
Exact Sciences, Tel-Aviv University, Tel-Aviv, 69978, Israel}

\date{\today}

\wideabs{

\maketitle

\begin{abstract}

Discrete breathers (nonlinear localised modes) have been shown to
exist in various nonlinear Hamiltonian lattice systems. In the
present paper we study the dynamics of classical spins interacting
via Heisenberg exchange on spatial $d$-dimensional lattices (with and
without
the presence of single-ion anisotropy). We show that
discrete breathers exist for cases when the continuum theory does
not allow for their presence (easy-axis ferromagnets with
anisotropic exchange and easy-plane ferromagnets). We prove the
existence of localised excitations using the implicit function
theorem and obtain necessary conditions for their existence. The
most interesting case is the easy-plane one which yields
excitations with locally tilted magnetisation. There is no
continuum analogue for such a solution and there exists an energy
threshold for it, which we have estimated analytically. We support
our analytical results with numerical high-precision computations,
including also a stability analysis for the excitations.

\end{abstract}


\newpage

}

%
\section{Introduction}

The phenomenon of dynamical localisation has been a subject of
intense theoretical research. It is well known that classical
Hamiltonian lattices possess periodic in time and
localised in space solutions called discrete breathers or
intrinsic localised modes. A recent explosion of interest to
discrete breathers has happened due to the fact that they may
exist in lattice models of interacting {\em identical} particles.
Breathers in continuum models (like, for example, the well known
sine-Gordon equation) exist only due to the  high symmetry
of the system and are therefore structurally unstable. Discrete
breathers are generic solutions of nonlinear lattice equations.
Their existence is based on the fact that all possible resonances
of multiples of their frequency with the bounded {\sl band of
small amplitude plane waves} (BSAPW) above the classical ground
state can be avoided. So far discrete breathers have been proven
to be generic solutions in both Hamiltonian\cite{sfcrw98,sa97} and
dissipative\cite{rsmjas98} systems. Several cases of experimental
observation  of discrete breathers have been reported [in
Josephson junction arrays\cite{baufz99-tmo99}, arrays of weakly
coupled waveguides \cite{hseys98-mpaes99}, low dimensional
crystals \cite{sblssbws99}, and biological systems
(myoglobin)\cite{xmhaprl00}].

Due to its spatial periodicity, lattices of interacting spins are
ideal systems to observe discrete breathers as well. Here we will
concentrate on large spins which may be described classically.
Nonlinear waves in magnetic systems have been extensively studied
during the last three decades\cite{kik90pr,mf84pr}. The results of
these studies provide a lot of information about the properties of
solitary waves (particularly, breathers) in magnets, since it is
possible in many cases to obtain explicit solutions to them.
However, neglecting discreteness effects may lead to loosing
important features of the nonlinear wave dynamics. For instance,
since only high symmetry continuous systems possess breather
solutions, the area of the potentially interesting models is
artificially reduced. Another drawback of continuous systems is
that the consideration of nontopological localised excitations is
typically restricted to space dimension one.

In the last decade, a number of papers has appeared where
localised modes in magnets were treated as essentially discrete
objects\cite{takeno91,ls99pr} (also an attempt of experimental
observation of discrete breathers in antiferromagnets has been
made\cite{ses99}). However, no rigorous existence proofs have been
given, and only the most simplest cases (from the point of view of
symmetries) have been considered. Preserving these symmetries one
can continue those solutions to the space continuous limit.

The aim of this work is to present breather excitations for spin
lattices in cases where the symmetries will not allow for a
similar mode construction in space continuous cases. Also we will
not restrict the consideration to one-dimensional systems. Below
we present a rigorous existence proof for discrete breathers in
magnetic systems using the anti-continuum limit. With the help of
this proof and of a Newton iteration method, we show the existence
of discrete breathers in ferromagnetic lattices with anisotropic
exchange interaction. We also
consider an easy-plane ferromagnet and find a new type of discrete
breathers, with several spins precessing around the hard
(single ion anisotropy) axis, while all the others precess around
an axis which lies in the easy plane. Note that only the simplest case
of monochromatic in
time breathers has been investigated in most of previous papers. Such a
situation simplifies considerably the treatment of the system, and
important families of solutions can be lost. Our studies do not
depend in any way on the number of higher harmonics in the time
evolution of a breather.

This paper is organised as follows. The next section presents the
model Hamiltonian and the equations of motion. In Sec. III, we
consider an easy-axis ferromagnet, discuss the implementation of
the anti-continuum limit, and give a rigorous existence proof for
discrete breathers. In Sec. IV, we study an easy-plane
ferromagnet. Then Sec. V presents a study of a two-dimensional
lattice with easy-plane anisotropy. Discussions and conclusions
are given in Sec. VI.

%
\section{Hamiltonian and equations of motion}

We consider a lattice of classical spins described by the
Hamiltonian with Heisenberg XYZ exchange interaction and
single-ion anisotropy
\begin{equation}
H=-\frac{1}{2}\sum_{{\bf n \neq n'}} \sum_{\alpha=(x,y,z)}
J_{\alpha}^{\bf n n'} S_{{\bf n}}^{\alpha} S_{{\bf n'}}^{\alpha}
-D \sum_{{\bf n}} {S_{{\bf n}}^z}^2~. \label{1}
\end{equation}
Here $S_n^x,S_n^y,S_n^z$ are the ${\bf n}$th spin components
(${\bf n}$ labels a lattice site) which satisfy the normalisation
condition
\begin{equation}
{S_{\bf n}^x}^2+{S_{\bf n}^y}^2+{S_{\bf n}^z}^2=S^2~.
\label{2}
\end{equation}
For simplicity, the total spin magnitude can be normalised to
unity: $S=1$. The constants $J_x,J_y,J_z>0$ are the exchange
integrals and $D$ is the on-site anisotropy constant.

The equations of motion for the spin components in the
one-dimensional spin chain with nearest-neighbour interactions are
the well known Landau-Lifshitz equations:
\begin{eqnarray}
{\dot S}_n^x&=&
\nonumber
\frac{1}{2}\left [J_y S_n^z \left (S_{n-1}^y+S_{n+1}^y\right )-
J_z S_n^y \left ( S_{n-1}^z+S_{n+1}^z \right)\right ] \\
\nonumber &-&2D S_n^y S_n^z ~,\\
{\dot S}_n^y&=&
\nonumber
\frac{1}{2} \left [J_z S_n^x \left (S_{n-1}^z+S_{n+1}^z \right )-
J_x S_n^z \left ( S_{n-1}^x+S_{n+1}^x \right) \right ] \\
\label{3} &+&2D S_n^x S_n^z ~,\\
{\dot S}_n^z&=&
\nonumber
\frac{1}{2} \left [J_x S_n^y \left (S_{n-1}^x+S_{n+1}^x \right )-
J_y S_n^x \left ( S_{n-1}^y+S_{n+1}^y \right) \right ] .
\end{eqnarray}
Generalisation to higher lattice dimensions is straightforward.

%
%
\section { Easy-axis ferromagnet}
%

First, we consider spin lattices with a ground state that
corresponds to all spins directed along a given axis (we assume
this axis to be the $Z$-axis). This can be achieved either by
introducing a strong exchange anisotropy ($J_x,J_y <<J_z$), or by
introducing an on-site anisotropy term $D>0$. Before we study
breather solutions of Eqs. (\ref{3}), let us study the dispersion
laws for linear spin waves.

\subsection {Dispersion laws}

First, we consider the easiest case, a ferromagnetic chain without
ion anisotropy ($D=0$), but  with a strong exchange anisotropy: $0
\leq J_x,J_y << J_z$. In this case, the ground state is $S_n^z=
\pm 1, S_n^x=S_n^y=0$. Linearising the equations of motion around
one of these ground states, e.g.,  $S_n^x = \delta_x \sin
{(qn-\omega t)}$, $S_n^y = \delta_y \cos {(qn-\omega t)}$, $S_n^z
= const  \approx 1$, we obtain
\begin{equation}
\omega_L^2(q)=(J_z-J_x \cos {q})(J_z- J_y \cos {q})~.
\end{equation}
This dispersion law is shown in Fig. 1, with the edges of the linear
band $\omega_L(q)$ being given by
\begin{equation}
\omega_0^2=(J_z-J_x)(J_z-J_y),~\omega_{\pi}^2=(J_x+J_z)(J_y+J_z)~.
\label{5}
\end{equation}
\begin{figure}[htb]
\vspace{1.0pt}
\centerline{\psfig{file=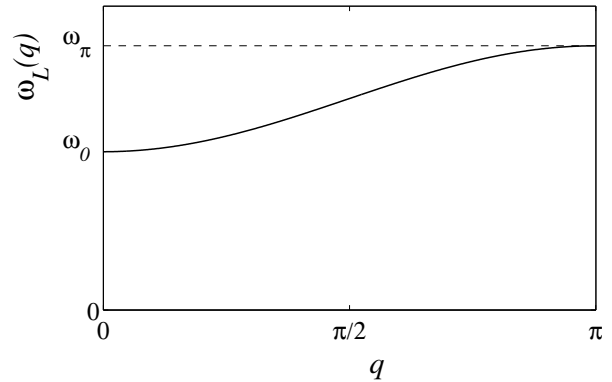,width=3.2in,angle=0}}
\vspace{4.5pt}
\label{fig1}
\caption{
Dispersion law for the ferromagnetic chain with strong exchange
anisotropy.}
\end{figure}

Let us explain how we can use the information about dispersion
relations in order to formulate expectations about existence or
nonexistence of breathers. It has been shown in \cite{f96pd} that
breather solutions of the full nonlinear equations bifurcate from
certain nonlinear plane wave solutions. These specific plane wave
solutions are time periodic and reduce to {\sl band edge plane
waves} (BEPW) in the limit of small amplitudes, i.e., to linear
waves with $\omega_L(q_{BEPW})$ being an extremum of $\omega_L(q)$. A
necessary prerequisite of the existence of breathers is that
the frequency of these BEPW's is detuning {\sl away} from the
linear band $\omega_L(q)$ with increasing amplitude or energy
density of the wave.

Let us consider an excitation which corresponds to the value of
the wave number $q=0$ from above. This is a space-homogeneous
excitation $S_n^{\alpha}=S^\alpha$ and the Landau-Lifshitz
equations yield
\begin{eqnarray}
\nonumber {\dot S}^x&=&(J_y-J_x) S^yS^z~, \\
 {\dot S}^y&=&(J_z-J_x) S^x S^z ~,\\
\nonumber {\dot S}^z&=&(J_x-J_y) S^xS^y ~.
\end{eqnarray}
As can be seen from these equations,
even for $J_x=J_y$ we have softening in the dispersion
law at
$q=0$. Then ${\dot S}^z=0$, and for $S^x =A \cos {\omega t}$ the spin
precession frequency
\begin{equation}
\omega^2=(1-A^2)(J_z-J_x)^2 < (J_z-J_x)^2 = \omega_0^2~,
\label{7}
\end{equation}
is below the lower edge of the linear band, which indicates an
occurrence of discrete breathers in the band gap. Note that for
the completely isotropic model $J_x=J_y=J_z \equiv J$, there is no
gap and consequently no breather solutions are to be expected.

It is easy to check that a similar analysis of the upper band edge
yields lowering of the BEPW frequency with increasing amplitude,
i.e. its frequency is {\sl attracted} by $\omega_L(q)$ instead of
being repelled. Consequently we do not expect breathers to
bifurcate from the upper band edge.

For $D\neq 0$ there are no qualitative changes. The ground state
of the chain remains to be the same, whereas the gap in the
dispersion law widens:
\begin{eqnarray}
\nonumber
\omega_L^2(q)&=&(J_z-J_x \cos {q})(J_z- J_y \cos {q})\\
&+&4D\left( J_z-\frac{J_x+J_y}{2}\cos {q} \right) + 4D^2~,
\label{8}
\end{eqnarray}
with
\begin{eqnarray}
\nonumber
\omega_0^2&=&(J_z-J_x)(J_z-J_y)+4D\left ( J_z-\frac{J_x+J_y}{2}\right
)+4D^2~,\\
&&\label{8a} \\
\omega_{\pi}^2&=&(J_x+J_z)(J_y+J_z)+4D\left (
J_z+\frac{J_x+J_y}{2}\right )+4D^2~.
\nonumber
\end{eqnarray}
Note that the band gap exists even in the case of isotropic
exchange (all $J_\alpha$'s are equal). Therefore, easy axis
anisotropy increases the chances of breather existence.

\subsection {Implementation of the anti-continuum limit}
\label{3B}

Now, following MacKay and Aubry\cite{ma94}, we apply the
anti-continuum (AC) limit to our system. The principle of the AC
limit consists in decoupling the lattice sites and exciting only
one or a small number of them, keeping all the other in the ground
state. Then, upon
switching on the interaction, the persistence of the localised
solution is shown. As a prerequisite for the successful existence
proof and continuation of the breather solution, the initial
`decoupled' periodic orbit must be anharmonic\cite{note1}, and the
breather frequency and all its multiples should not resonate with
the linear magnon band. In the case of strongly anisotropic
exchange ($J_x,J_y <<J_z$), the particular case of the AC limit
means simply setting $J_x=J_y=0$.


In this case the Z-component of each spin is conserved. The
solution of (\ref{3}) reduces to the precession of decoupled spins
around the $Z$ axis with frequencies which depend on the
realised value of the Z-components of nearest neighbour spins (due
to nonzero $J_z$):
\begin{equation}
S_n+iS_n^y= A_n e^{i(\omega_n t+\varphi_n)}~,
\label{10}
~~\dot{S}_n^z=0,
\end{equation}
where the precession frequency of the $n$th spin is given by
\begin{equation}
|\omega_n|=\frac{J_z}{2}(S_{n-1}^z+S_{n+1}^z)+2D
S_n^z,~A_n^2=1-{S_n^z}^2~.
\end{equation}

\subsubsection{Vanishing single-ion anisotropy ($D=0$)}

The initial choice of one precessing spin and all the others being
at rest
\begin{equation}
S_n^z= (...,1,1,1,S_0,1,1,1,...) \label{12}
\end{equation}
with $S_0< 1$ cannot be used to generate breathers because the
frequency of this solution $\omega=J_z$ resonates with the linear
band [more precisely, with the lower edge of the linear band
$\omega_0$, see Eq. (\ref{5})]. Therefore it cannot be continued
to the region of non-zero $J_x$ and $J_y$. A way out is simply to
excite three neighbouring spins:
\begin{equation}
 S_n^z=(...,1,1,1,S_1,S_0,S_1,1,1,1,...)~.
\label{13}
\end{equation}
Here $0<S_0<S_1<1$ and, since the precession frequency of all the
three central spins must be the same, $S_0=2S_1-1$. In this case,
the precession frequency $\omega=J_z S_1$, allowing for the
absence of resonances with the linear spectrum frequency
$\omega_0= J_z$, must satisfy the condition $k \omega \neq
\omega_0$.


\subsubsection{Additional single-ion anisotropy ($D>0$)}

In this case we may use the AC limit with the ansatz (\ref{12}).
The initial distribution of Z-spin components
are chosen as follows
\begin{equation}
\label{15} S_n^z=(...,1,1,1,S_0,1,1,1,...)~.
\end{equation}
Here one central spin is precessing with frequency $\omega=J_z+2DS^0$.
All other spins are
at rest. Small deviations from their equilibrium states yield precession
with frequencies
\begin{equation}
\omega_1= \omega_0+\frac{1}{2}(S_0-1) ~,~~ \omega_0=J_z (1+2D)~,
\label{16}
\end{equation}
distributed along the lattice in the following way
\begin{equation}
\omega_n= (...,\omega_0,\omega_0,\omega_1, \omega, \omega_1,
\omega_0, \omega_0,...)~.
\label{16a}
\end{equation}
Discrete
breathers can be continued from the AC limit if the following
non-resonance conditions are satisfied:
\begin{equation}
k \omega \neq \omega_0,~~ k\omega \neq \omega_1,~~ k \in Z~.
\label{17}
\end{equation}
Taking into account that $S_0=(\omega-J_z)/2D$ and substituting
it into the non-resonance condition $\omega \neq \omega_1$,
we get
\begin{eqnarray}
\nonumber
&&k \omega \neq \frac{J_z}{2}+2D-\frac{J_z^2}{4D}+\frac{J_z}{4D} \omega
 \\
\label{18}
&& \equiv -4D \left( \frac{J_z}{4D}-1\right)
\left (\frac{J_z}{4D}+\frac{1}{2} \right ) + \frac{J_z}{4D} \omega~.
\end{eqnarray}

Note that for $k=1$ the resonance will occur for any breather
frequency if $J_z=4D$. For this set of parameters, a breather
continuation from the AC limit is not possible for any frequency.

For this particular case we try another ansatz, namely the
even-parity case:
\begin{eqnarray}
\nonumber
 S_n^z=(...,1,1,1,S_0,S_0,1,1,1,...)~,\\
\omega_n=(...,\omega_0,\omega_0,\omega_1, \omega, \omega, \omega_1,
\omega_0, \omega_0,...)~
\label{19}
\end{eqnarray}
with $\omega_0$ and  $\omega_1$ being the same as in  Eq.
(\ref{16}), and
\begin{equation}
\omega=\frac{J_z}{2} \left ( 1+ S_0 \right ) +2 D S_0~.
\label{19a}
\end{equation}
Then, using the non-resonant condition (\ref{17}) which is valid
for this ansatz as well, we obtain
\begin{equation}
k\omega \neq
\omega_1=\frac{(J_z+4D)^2-J_z^2}{2(J_z+4D)}+\frac{J_z}{J_z+4D}\omega~.
\label{19b}
\end{equation}
It follows from this expression that the even-parity AC limit
allows for a continuation of the breather solution for all values
of $J_z$ and $D$.

\subsubsection{Isotropic exchange $J_x=J_y=J_z \equiv J$}

Here the ansatz (\ref{12}) can be used and the frequencies in the
AC limit will be distributed as $\omega_n= 2D S_n^z$. The
eigenfrequencies of the non-excited spins do not depend on the
values of the adjacent spins and equal $\omega_0$. Thus, the only
non-resonance condition to be fulfilled is $k\omega \neq \omega_0$.

%
%
\subsection {Existence proof for magnetic breathers}

Here we present a rigorous proof of the existence of discrete breathers
for the particular case of strongly anisotropic exchange and $D>0$.
\newtheorem{theorem}{Theorem}
\begin{theorem}
{\it If a periodic orbit of the Hamiltonian (\ref{1}) with a
frequency $\omega$ is non-resonant ($k\omega \neq \omega_{0,1}$,
$k\in Z$ and $\omega_L(q)$) and anharmonic\cite{note1}, then the
periodic orbit of the equations of motion (\ref{3}) at $\alpha
\equiv \{J_x,J_y\}=0$ given by the spin precession frequency
(\ref{15}) has a locally unique continuation as a periodic orbit
of the equations (\ref{3})  with the same period $T=2\pi/\omega$
for a sufficiently small $\alpha$.} \end{theorem}

{\em Proof}: Let $SL_T$ be the space of bounded infinite sequences
$z \equiv \{ z_n \}_{n \in Z}$ of triplets $z_n =
(S_n^x,S_n^y,S_n^z)$ of continuously differentiable functions of a
period $T=2 \pi/\omega$ with symmetry properties:
\begin{equation}
S_n^x(t)=S_n^x(-t),~S_n^y(t)=-S_n^y(-t),~S_n^z=S_n^z(-t)~.
\label{20}
\end{equation}
Then the size of oscillations on the $n$th site will be measured
by the following norm:
\begin{equation}
|z_n|=\sup {\left \{ |S_n^x(t)|,|S_n^y(t)|,|{\dot S}_n^x(t)|,
|{\dot S}_n^y(t)|; t \in R \right \}} .
\end{equation}
Next, the size of $z \in SL_T$ is given by
\begin{equation}
|z|=\sup { \left \{ |z_n|; n\in Z\right\}}
\end{equation}
and therefore $SL_T$ is a Banach space.

Consider now another Banach space $SM_T$ of bounded infinite
sequences $w \equiv \{w_n\}_{n \in Z}$ of triplets $w_n =
(M_n^x,M_n^y, M_n^z)$ of continuous functions of a period $T$ with
the following symmetry properties:
\begin{equation}
M_n^x(t)=-M_n^x(-t),~M_n^y(t)=M_n^y(-t),~M_n^z=-M_n^z(-t)~,
\end{equation}
and the norms
\begin{eqnarray}
\nonumber
|w_n|&=& \sup {\left \{ |M_n^x(t)|,|M_n^y(t)|,|M_n^z(t)-1|; t \in R
\right \} }, \\
|w|&=& \sup {\left \{ |w_n|; n \in Z \right \} }.
\end{eqnarray}
Define the mapping $ F$: $SL_T \rightarrow SM_T$
$$
F(z, \alpha)=w
$$
with $\alpha=\{J_x,J_y \}$, $z=\{S_n^x,S_n^y,S_n^z\}_{n \in Z}$,
 $w=\{M_n^x,M_n^y,M_n^z\}_{n \in Z}$, and
\begin{eqnarray}
M_n^x&=&
\nonumber
\frac{1}{2}\left [J_y S_n^z \left (S_{n-1}^y+S_{n+1}^y\right )-
J_z S_n^y \left ( S_{n-1}^z+S_{n+1}^z \right) \right ] \\
\nonumber &-&2D S_n^y S_n^z-{\dot S}_n^x~, \\
M_n^y&=&
\nonumber
\frac{1}{2}\left [J_z S_n^x \left (S_{n-1}^z+S_{n+1}^z \right )-
J_x S_n^z \left ( S_{n-1}^x+S_{n+1}^x \right) \right ] \\
\label{25} &+&2D S_n^x S_n^z-{\dot S}_n^y~, \\
M_n^z&=&
\nonumber
\frac{1}{2} \left [
J_x S_n^y \left (S_{n-1}^x+S_{n+1}^x \right )-
J_y S_n^x \left ( S_{n-1}^y+S_{n+1}^y \right) \right ]\\
\nonumber &-&{\dot S}_n^z ~.
\end{eqnarray}
The symmetric solutions of the equations of motion (\ref{3}) are
in one-to-one correspondence with zeroes of $F$, i.e.,
$F(z,\alpha)=0$, especially in the case when $\alpha=0$. Using the
implicit function theorem, we prove this solution to have a
locally unique continuation $z(\alpha)$ for sufficiently small
$\alpha$, such that $F[(z(\alpha),\alpha]=0$ provided $F \in C^1$
and its derivative with respect to $z$, ${\cal D}$, is invertible
at $\alpha=0$.

To show the invertibility of ${\cal D}F$,  we linearise the map
around the periodic orbit at $\alpha=0$
\begin{equation}
{\bf \delta  M} = {\cal D} F {\bf \delta S}
\label{25a}
\end{equation}
and show that it is invertible simultaneously in the following
three parts of the lattice: the central site $n=0$, its adjacent
sites $n=\pm 1$, and the remainder of the lattice. Invertibility
of ${\cal D} F$ is equivalent to the invertibility of the
corresponding matrix in (\ref{25a}).

(i) For $n \neq 0, \pm 1$ we have
\begin{eqnarray}
\nonumber \delta M_n^x&=&-\delta S_n^y \omega_0 - \delta {\dot S}_n^x~,
\\
\delta M_n^y&=&\delta S_n^x \omega_0 - \delta {\dot S}_n^y~, \\
\nonumber \delta M_n^z&=& - \delta {\dot S}_n^z~.
\end{eqnarray}
Since the functions $\delta M_n^\alpha$ and
$\delta S_n^\alpha$ are periodic in time,
we can expand them into Fourier series
\begin{equation}
\delta...=\sum_{k=-\infty}^{+\infty} \delta... (k) e^{ik\omega t}~.
\end{equation}
The time-symmetry requirements (\ref{20})
ensures $\delta...(k \geq 0)$ to be either purely real or imaginary,
so that
\begin{eqnarray}
\nonumber \delta M_n^x(-k)&=&- \delta M_n^x(k)~,~~\delta
M_n^y(-k)=\delta M_n^y(k)~,\\
\nonumber &&\delta M_n^z(-k)=- \delta M_n^z(k)~,\\
\delta S_n^x(-k)&=& \delta S_n^x(k)~,~~\delta S_n^y(-k)=- \delta
S_n^y(k)~,\\
\nonumber &&\delta S_n^z(-k)= \delta S_n^z(k)~.
\end{eqnarray}
Particularly, $\delta M_n^x(0)=\delta M_n^z(0)=\delta S_n^y(0)=0$.
As a result, for $k \geq 0$ we obtain
\begin{eqnarray}
\nonumber
\delta M_n^x(k)&=&- \omega_0 \delta S_n^y(k) - ik\omega \delta
{S}_n^x(k)~, \\
\delta M_n^y(k)&=& \omega_0 \delta S_n^x(k) - ik\omega \delta
S_n^y(k)~,\\
\nonumber
\delta M_n^z(k)&=& -ik\omega \delta {S}_n^z(k)~.
\end{eqnarray}
These equations appear to be decoupled with respect to $k$ and
they can be inverted if $k^2 \omega^2 \neq \omega_0^2$. The third
equation can be inverted for $k \neq 0$ if $\omega \neq 0$. For
$k=0$ the inversion is impossible, but this degeneracy can be
lifted by imposing the normalisation condition (\ref{2}). In fact,
the third equation can be dropped because $S_n^z$ is defined by
$S_n^x$ and $S_n^y$ with accuracy up to a sign.
 All that we have to check is the inequality
${S_n^x}^2+{S_n^y}^2-1 \leq 0$.

(ii) For $n=\pm1$ we act similarly to the previous case:
\begin{eqnarray}
\nonumber \delta M_1^x&=&-\delta S_1^y \omega_1 - \delta {\dot S}_1^x~,
\\
\delta M_1^y&=&\delta S_1^x \omega_1 - \delta {\dot S}_1^y~, \\
\nonumber \delta M_1^z&=& - \delta {\dot S}_1^z~.
\end{eqnarray}
A similar condition for invertibility can be found: $k^2 \omega^2 \neq
\omega_1^2$.

(iii) Case $n=0$.
Using the canonical coordinate $y$ and momentum $p$
\begin{equation}
S_0^x=\sqrt{1-y^2} \cos {p},~ S_0^y=\sqrt{1-y^2} \sin {p},~S_0^z=y,
\label{31}
\end{equation}
we obtain the following Hamilton equations:
\begin{equation}
{\dot y}=\frac{\partial H}{\partial p}~,~~{\dot p}=-\frac{\partial
H}{\partial y}~.
\label{31a}
\end{equation}
We define the pair $(u,v)$ instead of the set $(M_0^x,M_0^y,M_0^z)$,
with
\begin{eqnarray}
u=\frac{\partial H}{\partial p}- {\dot y}~,~~v=-{\dot p}-\frac{\partial
H}{\partial y}~.
\end{eqnarray}
The new function satisfies the following symmetries:
\begin{eqnarray}
\nonumber y(t)&=&y(-t),~p=\omega t + h(t),~h(t)=-h(-t),\\
y(t+T)&=&y(t)~,~~h(t)=h(t+T)~.
\end{eqnarray}
Here $h(t)$ is a periodic function in time.
Keeping in mind that $S_{-1}^z$ and $S_1^z$ are fixed by
variations of $S_{\pm 1}^{x,y}$, we obtain
\begin{eqnarray}
\nonumber
\delta u &=& 2 D \delta y - \delta {\dot p}= 2 D \delta y -
\delta {\dot h}~,\\
\delta v &=& - \delta {\dot y}~.
\end{eqnarray}
After considering the corresponding Fourier series with respect to
time, we find the following inversion conditions:
 $\omega \neq 0$ for $k\neq 0$ and  $D \neq 0$ for $k=0$.

Once the invertibility is shown, by the implicit function theorem, the
initial
solution $z(0)$ has a locally unique continuation $z(\alpha)$ in
$SL_{T}$, and thus the theorem has been proved. $\Box$

This result can be easily extended to lattices in  higher
dimensions, antiferromagnets, and systems with larger interaction
radius.

%
\subsection {A method of computation of discrete breathers and linear
stability  analysis}

For numerical simulations it is convenient to use stereographic
coordinates. The new coordinates incorporate the normalisation
condition and reduce the problem with three unknown real functions
per site to the problem with one unknown complex function per site:
\begin{equation}
\xi_n=\frac{S_n^x+iS_n^y}{1+S_n^z}~.
\end{equation}
The inverse transform is given by
\begin{equation}
S_n^x=\frac{\xi_n+\xi_n^*}{1+|\xi_n|^2},~
S_n^y=\frac{1}{i}\frac{\xi_n-\xi_n^*}{1+|\xi_n|^2},~
S_n^z=\frac{1-|\xi_n|^2}{1+|\xi_n|^2}~.
\end{equation}
In these new coordinates the Landau-Lifshitz equations take the form
\begin{eqnarray}
&&{\dot \xi}_n=\frac{1}{4i} \left [ (J_x+J_y)
\left ( \frac{\xi_{n-1}-\xi_n^2\xi_{n-1}^*}{1+|\xi_{n-1}|^2}+
\nonumber
 \frac{\xi_{n+1}-\xi_n^2\xi_{n+1}^*}{1+|\xi_{n+1}|^2} \right )\right.\\
 &&+ \left.
 (J_x-J_y)
\label{37}
 \left ( \frac{\xi_{n-1}^*-\xi_n^2\xi_{n-1}}{1+|\xi_{n-1}|^2}+
 \frac{\xi_{n+1}^*-\xi_n^2 \xi_{n+1}}{1+|\xi_{n+1}|^2}
 \right ) \right .\\
&&- \left .
2 J_z \xi_n \left ( \frac{1-|\xi_{n-1}|^2}{1+|\xi_{n-1}|^2}+
 \frac{1-|\xi_{n+1}|^2}{1+|\xi_{n+1}|^2}\right )
 -8D \xi_n \frac{1-|\xi_n|^2}{1+|\xi_n|^2}\right ].
\nonumber
\end{eqnarray}
The computation of discrete breathers is done using the Newton
map\cite{sfcrw98}.

The linear stability analysis of discrete breathers is performed
by linearising Eqs. (\ref{37}): $\xi_n(t) = \xi_n^{(0)}(t) +
\epsilon_n(t)$ around the breather periodic orbit, and solving
afterwards the eigenvalue problem
\begin{equation}
\left ( \begin{array}{c} \mbox{Re} ~\epsilon_n(T) \\ \mbox{Im}
~\epsilon_n(T)
\end{array} \right )=
\hat {{\cal M}}  \left ( \begin{array}{c} \mbox{Re}~ \epsilon_n(0) \\
\mbox{Im}~ \epsilon_n(0) \end{array} \right )~.
\end{equation}
If the eigenvalues $\Lambda$ of the Floquet matrix $\hat {{\cal M}}$ are

found to be located on the unit circle of the complex plane, then
according to the Floquet theorem, the periodic orbit is stable,
otherwise it is unstable.

%
\subsection {Breather solutions in an easy-axis ferromagnet}

Breathers in magnetic lattices with an easy-axis anisotropy can be
viewed as localised spin excitations with the spins precessing
around one of the ground states of the system (which we have chosen
$S^z=1$ at the beginning of the paper), so that the effective
radius of this precession decreases to zero at $n \rightarrow \pm
\infty$. The case of an isotropic exchange in $XY$ ($J_x=J_y$) is
the simplest one because the $S^z$ component is conserved in the
solution, and therefore the separation of the time and the space
variables
\begin{equation}
S_n^+=S_n^x+iS_n^y= A_n e^{i\omega t}
\label{38a}
\end{equation}
is possible in the Landau-Lifshitz equation [this can easily be
seen also from Eq. (\ref{37})]. The precession amplitudes $A_n$ do
not depend on time and the Landau-Lifshitz equations (\ref{3}) are
reduced to a set of algebraic equations which can be solved by a
simple iteration procedure. This is true both in the case of
strong exchange anisotropy $J_x=J_y \ll J_z$ and also when the
exchange is isotropic and an on-site anisotropy $D>0$ is present.

The breather existence in the $J_{x,y}-D$ plane is governed by the
non-resonance conditions given in Subsec. \ref{3B} for one single
harmonics $k=1$. With the growth of $J$ the  breather frequency
may hit the linear spectrum which marks the boundary of breather
existence on this plane. The nature of the other non-resonance
condition (\ref{18}) is different, e.g., we cannot continue a
breather solution for small  $J=J_{x,y}$ when $J_z=4D$, however
breathers exist for larger values of  $J$. Note, that in the case
of $J_x=J_y$ discrete breathers have a continuum equivalent which
is a breather solution of the integrable nonlinear Schr\"odinger
equation. The reason is that the $XY$ exchange symmetry allows one
to find solutions which are monochromatic in time (\ref{38a}). As
long as the linear band provides a gap and the nonlinearity allows
for pushing the breather frequency into the gap, localised
excitations may be found regardless of the degree of discreteness
of the system, which can be characterised by the ratio of the gap
to the bandwidth of the spin wave spectrum.

\begin{figure}[htb]
\vspace{2.0pt}
\centerline{\psfig{file=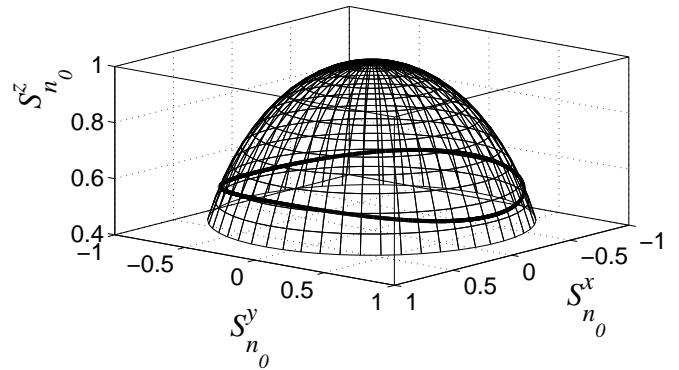,width=3.5in,angle=0}}
\vspace{5.5pt}
\caption{Dynamics of the central spin with site
number $n_0=11$ in the chain of $N=21$ spins with $J_x=0.1$,
$J_y=0.23$, $J_z=1$, $D=0$.}
\label{fig2}
\end{figure}

\begin{figure}[htb]
\vspace{0.20pt}
\centerline{\psfig{file=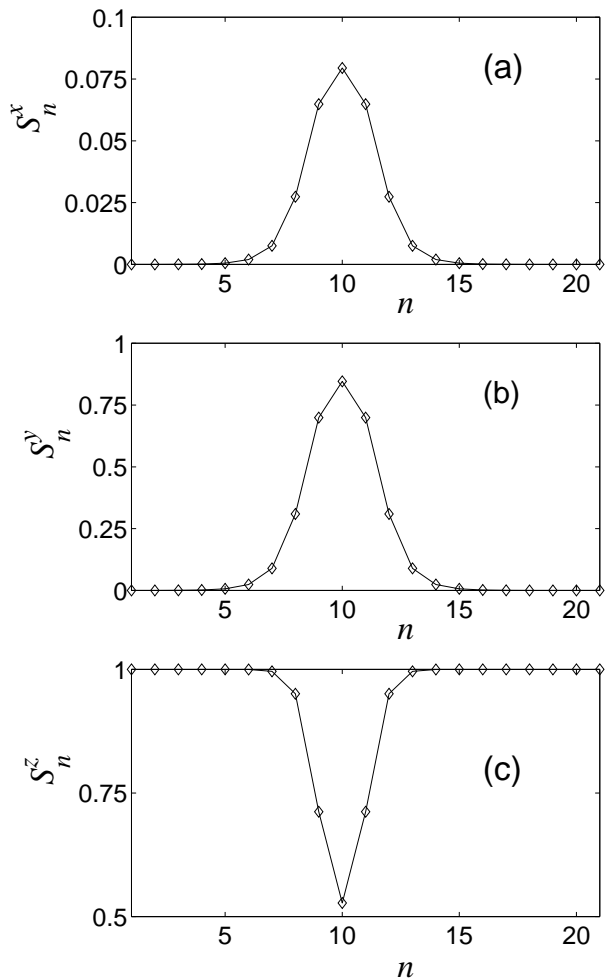,width=3.20in,angle=0}}
\vspace{2.0pt}
\caption{Discrete breather profile at a fixed
instant of time for the parameters described in Fig. \ref{fig2}.}
\label{fig3}
\end{figure}

Now we consider a rhombic chain with $J_x \neq J_y < J_z$.
Breaking isotropy in the $XY$ plane implies that $S^z$ is not
conserved in the solution anymore, and according to the
Landau-Lifshitz equations, it is impossible to represent the
breather solution in the form (\ref{38a}). This implies that
breathers will have an infinite number of harmonics in time, and
consequently
the spin dynamics is more complicated. Each spin now
draws an `elliptic' trajectory on the unit sphere, elongated
towards the larger component of $J_x$ or $J_y$. Fig. \ref{fig2}
%
%
\noindent shows such a dynamics
of the central spin $n_0=11$ of the breather in a chain consisting
of $N=21$ spins. The breather profile at some instant of time is
shown in Fig. \ref{fig3}.
%
%

Due to the broken symmetry in the $XY$ plane we are not able to
find  breathers in the corresponding continuum problem. The reason
simply is that the linear band of a continuum equation may still
have a gap but will be unbounded from above. Consequently there
will be unavoidable resonances of higher harmonics of a breather
with the linear band causing in general nonexistence of the
solution itself. Here we have a nontrivial case where the
discreteness of the lattice provides the necessary support for
breathers which is missing in the continuum case. The computation
of breather periodic orbits in this case cannot be reduced to
solving a system of algebraic equations and we have to work in the
full phase space by using, e.g., a generalised Newton
map\cite{sfcrw98}.

\begin{figure}[htb]
\vspace{0.40pt}
\centerline{\psfig{file=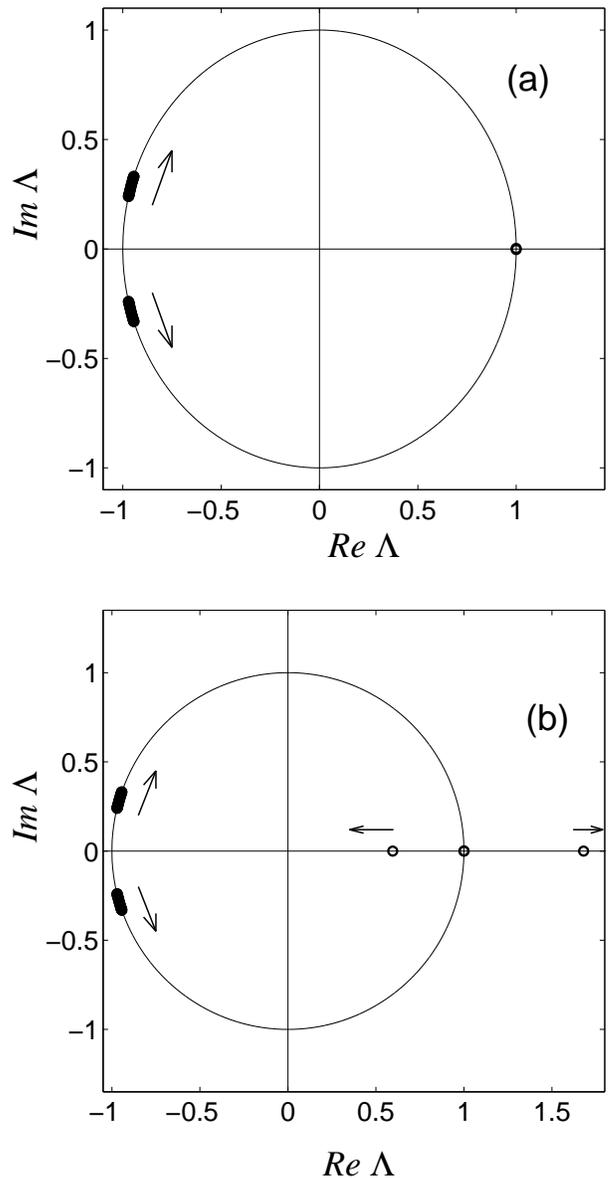,width=3.20in,angle=0}}
\vspace{3.pt}
\caption{
Eigenvalues $\{ \Lambda \}$ of the Floquet matrix for (a) site-centred
breather
and (b)  bond-centred breather in the easy-axis chain of $N=32$ spins
with $D=1$, $J=0.01$, $\omega=1.3$. Arrows show the direction
of motion of eigenvalues when $J$ is increased.}
\label{fig4}
\end{figure}

Let us briefly discuss the stability of the obtained breather
solutions. Previous stability studies \cite{ma96n} have shown that
the stability depends on the breather parity (i.e., its spatial
symmetry). Our Floquet analysis of the eigenvalues of the
stability matrix confirms these findings. We obtain that the
site-centred breathers (continued from the one-site breather) are
stable in the limit of small exchange (see Fig. \ref{fig4}a)
%
%
\noindent while the bond-centred breathers [continued from the
two-site breathers, see ansatz Eq. (\ref{19})] are unstable
arbitrarily close to the AC limit with the unstable eigenvalue
being located on the positive half or the real axis outside of the
unit circle (see Fig. \ref{fig4}b).

%
%
\section {Easy-plane ferromagnet}

In the case of an easy-plane anisotropy we choose $D<0$ and
$J_x=J_y=J_z \equiv J$. The ground state of the system, without
loss of generality, can be assumed to be
\begin{equation}
S_n^x=1, ~~S_n^y=S_n^z=0~.
\label{40}
\end{equation}
Note that the ground state is degenerate, so that the spins can be
oriented arbitrarily in the $XY$ plane, but they must stay parallel to
each other .

%
%
\subsection {Linear dispersion law and the anti-continuum limit}

Linearising the equations of motion in the vicinity of the
ground-state (\ref{40}) we obtain the following dispersion
law:
\begin{equation}
\omega^2 (q)=J^2 (1- \cos {q})^2+ 2J|D|(1-\cos {q})~.
\label{39}
\end{equation}
This is an `acoustic'-type dispersion law with
\begin{equation}
\omega_0^2=\omega^2(0)=0~,~~\omega_{\pi}^2=\omega^2(\pi)=4J(J+|D|)~,
\label{39a}
\end{equation}
and therefore the breather frequencies in this case
should lie above the linear band.

The implementation of the AC limit can be achieved by setting
$J=0$ and exciting one or several spins, so that they would start to
precess
around the hard axis with a frequency $\omega=2|D|S_0$, where
$S_0$ is the $z$ projection of the spin. If the non-resonance
condition $\omega \neq \omega_{\pi}=0$ is satisfied, breather
solutions can be continued.

Breathers do not exist in the continuum limit with easy-plane
anisotropy\cite{ikb79,bk80}. The reason is again that the
corresponding linear band is gapless and unbounded, so that it
covers the whole real axis. Correspondingly, there is no place for
a frequency of a localised excitation on the real axis which does
not resonate with the linear band. An essentially discrete model
\cite{wmb95} has been studied only in the case of a strong
magnetic field directed along the hard axis. In this case the hard
axis effectively becomes an easy axis and, as a result, the spins
precess around the $Z$ axis with a constant $S^z$ component. In
this case, a separation of variables (\ref{38a}) is possible
which simplifies the treatment of the system.

%
%
\subsection {Breather solutions of the easy-plane ferromagnet}

\begin{figure}[htb]
\vspace{2.0pt}
\centerline{\psfig{file=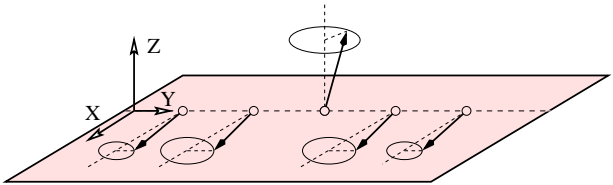,width=0.95in,angle=-90}}
\vspace{9.5pt}
\caption{Schematic representation of a discrete
breather with one `out-of-plane' spin in the easy-plane magnet.}
\label{fig5}
\end{figure}

\begin{figure}[htb]
\vspace{2.0pt}
\centerline{\psfig{file=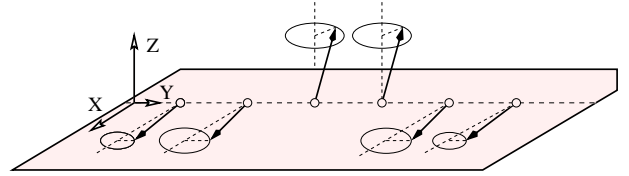,width=0.90in,angle=-90}}
\vspace{9.5pt}
\caption{ Schematic representation of the discrete
breather with two parallel `out-of-plane' spins in the easy-plane
ferromagnet.}
\label{fig6}
\end{figure}

As stated above, we do not apply an external magnetic field, and
therefore we do not change the ground state. As in the previous
case of easy-axis, we compute breather periodic orbits from the AC
limit using the generalised Newton method. As a result we obtain
solutions, schematically described in Fig. \ref{fig5} (for one
`out-of-plane' spin)
%
%
\noindent or for two parallel precessing
`out-of-plane' spins, as shown in Fig. \ref{fig6}.
%
%
For non-zero $J$ previously non-excited spins start to precess with
small amplitudes around the $X$-axis while
the plane of precession of the `out-of-plane' spin is
no longer parallel to the easy plane, but is slightly tilted.
Breathers with more than two precessing spins can be also created.

Depending on its frequency, the breather width changes. When the
frequency approaches the upper edge of the linear band, the
breather becomes more delocalised. However, this does not
qualitatively influence its core structure, i.e., the effective
precessing axis of the central spin is not continuously tilted
towards the $X$ axis upon lowering the breather frequency down to
the linear band edge. The central spin dynamics can be viewed as a
periodic (closed) orbit of a point confined to the unit sphere.
Let the $XY$ plane be the equatorial one. Then for large breather
frequencies the point performs small circles around the north (or
south) pole. Lowering the breather frequency does not change the
fact that the loop still encircles the $Z$-axis. Thus the breather
solution can not be deformed into a slightly perturbed and weakly
localised BEPW. This makes clear that the easy-plane ferromagnet
lattice supports breather solutions with a local magnetisation
tilt which have no analog in a continuum theory.

\begin{figure}[htb]
\vspace{1.0pt}
\centerline{\psfig{file=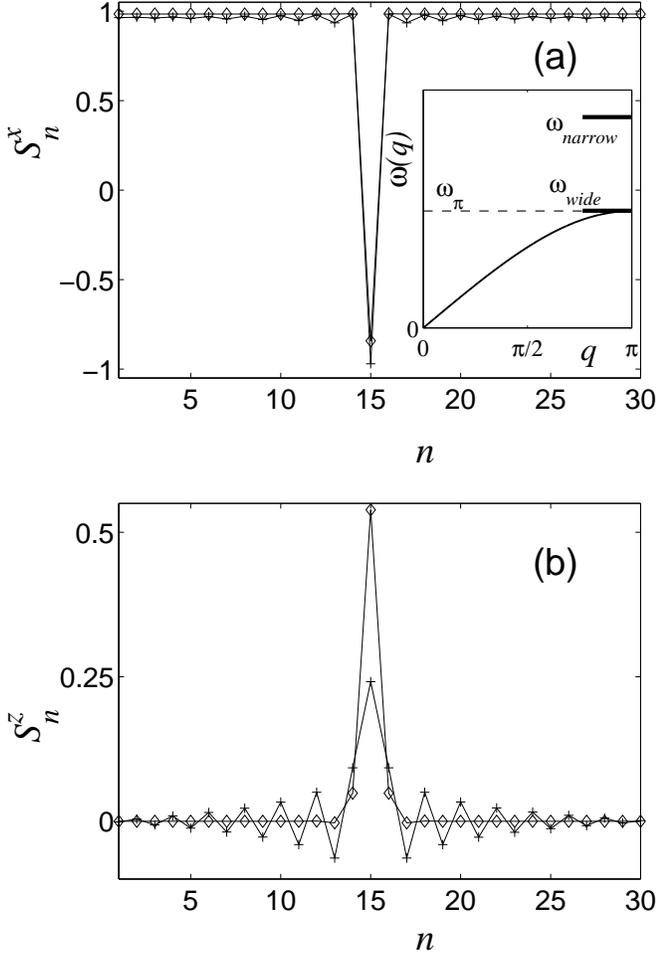,width=3.50in,angle=0}}
\vspace{4.pt} \caption{Discrete breather profile $J=0.1$, $D=-1$,
$\omega_{narrow}=1.1967$ (diamonds) and $\omega_{wide}=0.6649$
(crosses). Inset shows the linear dispersion law and the location
of breather frequencies.}
\label{fig7}
\end{figure}

The situation is illustrated by Fig. \ref{fig7},
%
%
\noindent where the profiles of two breathers are represented: one
corresponds to the frequency $\omega_{wide}=0.6649~$, which is
very close to the upper edge of the linear band
$\omega_{\pi}=0.6633$ and the other one has the frequency
$\omega_{narrow}=1.1967$, which is far above the linear band (see
the inset to Fig. \ref{fig7}).

\begin{figure}[htb]
\vspace{1.0pt}
\centerline{\psfig{file=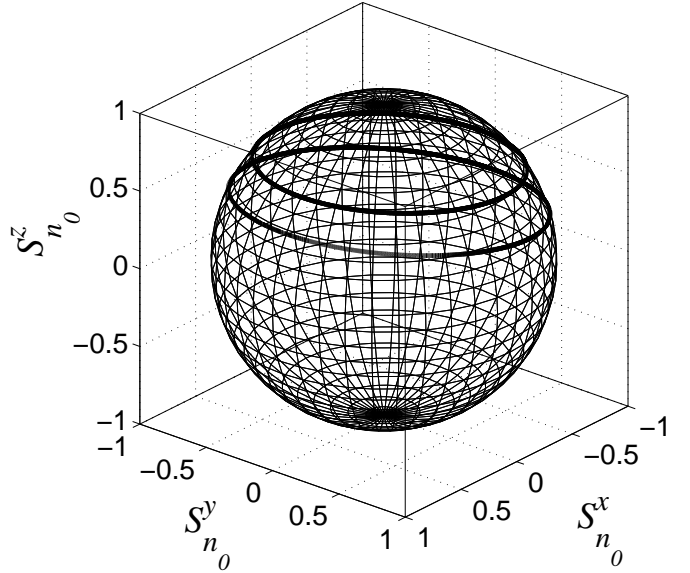,width=3.50in,angle=0}}
\vspace{4.pt}
\caption{Dynamics of the central spin with site
number $n_0=15$ in the chain, described in Fig. \ref{fig7}.}
\label{fig8}
\end{figure}

\begin{figure}[htb]
\vspace{1.0pt}
\centerline{\psfig{file=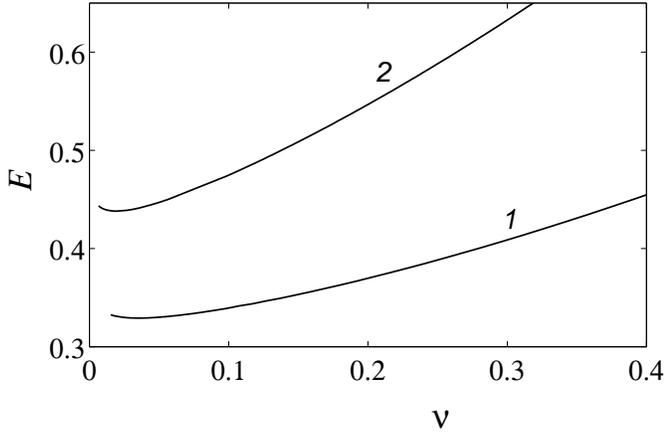,width=3.50in,angle=0}}
\vspace{4.pt}
\caption{
Normalised energy $E=H+JN/2$  as a function
of the detuning frequency $\nu = \omega-\omega_{\pi}$ for
a discrete breather with one out-of plane spin
(curve 1) and with two out-of-plane spins (curve 2) with $J=0.1$.
The size of the
system is $N=30$ spins.}
\label{fig9}
\end{figure}

The first solution is more delocalised, which can be seen from
Fig. \ref{fig7}. However, the central spin still precesses in a way
similar to the `narrow' breather, i.e. it encircles the
north pole on some lower latitude as compared to the narrow
breather (see Fig. \ref{fig8}, two curves on the unit sphere
correspond to two breather periodic orbits, discussed above).
%
%
Hence, even when very close to the linear band, our
breathers have a structure, which has no analogue in the continuum
case. Moreover, we have investigated the dependence of the breather
energy on the breather frequency (see Fig. \ref{fig9}). We observe
that there exists an energy threshold since the breather energy
attains a non-zero minimum when its frequency is still not equal
to the edge of the linear spin wave spectrum. Note that for lattices of
interacting scalar degrees of freedom discrete breathers have
typically zero lower energy bounds in spatial dimension $d=1$ and
become nonzero only for $d=2,3$ \cite{fkm97prl}. The reason for
the appearance of a lower nonzero bound in the present case is due
to the already mentioned fact that the breather of the easy plane
ferromagnet system does not deform into a perturbed band edge
magnon wave. Instead the central spin(s) is precessing around the
$Z$ axis. This topological difference is the reason for the
appearance of nonzero lower energy bounds. Such energy thresholds
may be very important as they show up in contributions to
thermodynamic quantities which depend exponentially on
temperature. To eliminate possible size effects, we repeated
calculations for
Fig. \ref{fig9} for a chain with $N=50$ spins. The difference
between the curves was negligibly small.
%
%

Energy thresholds can be
estimated analytically in the limit of small exchange $J$.
Ignoring displacements of all in-plane spins we obtain the
threshold energy for a breather  with $M$ out-of-plane precessing spins,
normalised
to the ground state:
\begin{equation}
E(\omega) \approx 2 M |D|{S_0^z}^2~,
\label{thresh}
\end{equation}
where $S_0^z=\omega/2|D|$ is $Z$-component of the precessing spin in the

AC limit [see Eq. (\ref{3})]. The factor $2$ comes from the fact
that we should take into account the contribution of the breather tails.

Equation (\ref{thresh}) can be obtained using the following
argumentation. For
large values of $\nu$ in Fig. \ref{fig9} the main contribution to the
breather
energy comes from the $M$ `out-of-plane' precessing spins, because the
tail
amplitudes of the breather are small (see, for example, Fig.
\ref{fig7}).
For $\nu \rightarrow 0$ the energy contribution from the tails is
actually
diverging. Thus the height of the minima of the curves in Fig.
\ref{fig9} can be estimated
as two times the contribution coming from the central spins.
Substitution into the above formula
for the band edge frequency $\omega_{\pi}$ yields
$\bar {E} = E(\omega_{\pi}) \approx 2MJ+ O(J^2)$.
]~,
For the case, considered in Fig. \ref{fig7}, for $M=1$ (one precessing
spin)
the analytic result
yields $\bar {E} \approx 0.2$ while numerics give the value
$0.33$.
In the case of two precessing spins ($M=2$)
the numerical result yields $\bar {E} \approx 0.44$
while analytical estimate predicts the value $0.4$.


Increase of the frequency leads to a decrease of the precession
radius of the central spin. In the AC limit, an upper bound for
the breather frequency is defined by $\omega=2D$,  which
corresponds to the central (precessing) spin being parallel to the
$Z$ axis. This bound continues to exist when exchange is switched
on. After reaching this frequency threshold, the breather becomes
a stationary (time-independent) solution. The existence of such a
solution has been verified numerically by solving the
time-independent Landau-Lifshitz equations.

\subsection{Stability of breather solutions and their asymptotic
properties}

We have investigated the stability of our solutions with the help
of the Floquet analysis (for details see Subsec. IIID) and with direct
Runge-Kutta simulations.
It appears that for small $J$'s breathers with one precessing spin
(see Fig. \ref{fig5}) are unstable, while
 a stable configuration which corresponds to two parallel
precessing spins (see Fig. \ref{fig6}) is stable. Note that similar
results have been obtained for the FPU-type lattices\cite{sps92prb}.

\begin{figure}[htb]
\vspace{2.0pt}
\centerline{\psfig{file=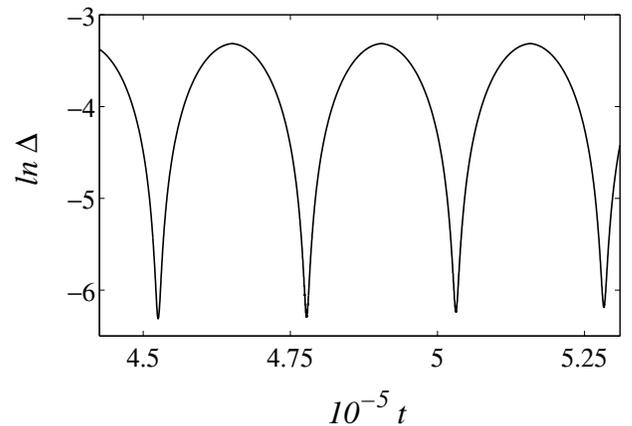,width=3.250in,angle=0}}
\vspace{5pt}
\caption{Time dependence of the effective error
$\Delta$ for the breather solution with $J=0.1$, $D=-1$ and
oscillation period $T=4.4248$. }
\label{fig10}
\end{figure}

Stability tests also
included the following numerical experiment. A periodic breather
orbit $\{{S_n^x(t)}^{(0)},{S_n^y(t)}^{(0)},{S_n^z(t)}^{(0)} \}$ is
perturbed by deviating one of the central spins $S_n^z(0)^{(0)}
\rightarrow S_n^z(0)^{(0)}+\varepsilon$ and simulating the
equations of motion (\ref{3}). The error function
\begin{equation}
\Delta (t)= \min_{\tau \in [0,T]}\sqrt {\sum_{n=1}^N
\sum_{\alpha=(x,y,z)}
\left [ S_n^{\alpha}(t) - {S_n^{\alpha}}^{(0)}(\tau) \right ]^2 }
\label{46}
\end{equation}
was calculated on each breather oscillation period $T$. In Fig.
\ref{fig10}, such a function (with $\varepsilon=-0.0025$,
in a
chain consisting of $N=1000$ spins with periodic
boundary conditions) is shown for the breather
solution with the two in-phase precessing spins
(see Fig. \ref{fig6}).
%
%
The error function is bounded during a significant period of time
(more than 10000 breather oscillation periods) and the breather
structure remains preserved. Similar experiments have been
performed for other types of breathers. They yield similar
results. This demonstrates the stability of the discussed excitation.

For a better understanding of the internal breather dynamics,
the data will be represented using the Fourier expansion of the breather

periodic orbit
\begin{eqnarray}
\nonumber S_n^{\alpha}(t)&=& C_0^{\alpha}(\omega;n)+
\sum_{k=1}^{\infty} \left [ A_k^{\alpha}(\omega;n) \cos {k \omega
t}+\right. \\
&+& \left . B_k^{\alpha}(\omega;n) \sin {k \omega t} \right ]~,
~~\alpha=x,y,z~.
\label{41}
\end{eqnarray}
We plot the space configuration  of $ C_0$ and
$C_n= \sqrt {A_n^2+B_n^2}$.

\begin{figure}[htb]
\vspace{2.0pt}
\centerline{\psfig{file=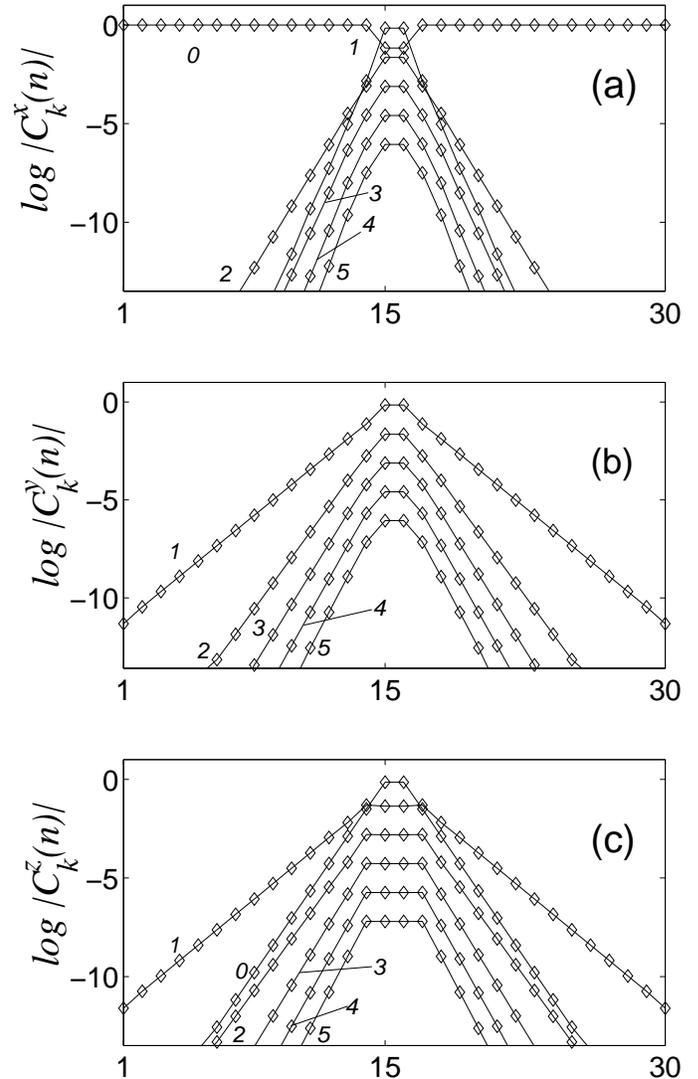,width=3.60in,angle=0}}
\vspace{9.5pt}
\caption{Spatial dependence of the Fourier
components of the discrete breather of the type depicted in Fig.
\ref{fig8} for $J=0.18$, $D=-1$, $\omega=1.42$. Numbers on the
panels represent the order of the harmonic $k$ [see Eq. (\ref{41})
for explanation]. }
\label{fig11}
\end{figure}

The `logarithmic' profile of such a solution is given in Fig.
\ref{fig11}. We have plotted the space dependence of its Fourier
harmonics (from the zeroth to the fifth one) for one particular
stable solution.

Let us analyse the behaviour and the exponential spatial decay of these
harmonics.  As can be seen from Fig. \ref{fig11}, the zeroth
(static) component is present. According to this figure, the
zeroth component decays exponentially in space. This seems to be
surprising, since the corresponding zero frequency resonates with
the bottom of the acoustic-type linear band [see Eqs.
(\ref{39})-(\ref{39a})].

%
%
To understand the results in Fig. \ref{fig11} we linearise the
equations of motion (\ref{3}) around the ground state (\ref{40})
in the breather tails. We obtain the following equations for $S^y$
and $S^z$ components ($S^x$ is assumed to be equal $1$ with higher
than linear corrections):

\begin{eqnarray}
\nonumber
&&-\frac{J}{2|D|} (S_{n+1}^z-2 S_n^z+S_{n-1}^z) + 2S_n^z=0~,\\
\label{43}
&& S_{n+1}^y-2 S_n^y+S_{n-1}^y=0~.
\end{eqnarray}

The numerical results suggest that the static $S^y_n$ component is
zero. This satisfies the second equation in (\ref{43}). The first
equation in (\ref{43}) allows for an exponential decay of the
static $S^z_n$ component. Its decay can be characterised by the
value $\lambda_0^z$ if $C_0^z(\omega;n) \sim \exp {(-\lambda_0^z
|n|)}~,~|n| \rightarrow \infty ~,~\lambda_0^z>0 $. The substitution of
this ansatz into (\ref{39}) yields

\begin{equation}
\lambda_0^z= \ln \left [1+\frac{2|D|}{J}+
\sqrt{\left (1+\frac{2|D|}{J}\right )^2-1} \right ]~.
\label{44}
\end{equation}

The spatial decay of all the other (non-zero) harmonics of the
breather solution can be obtained from the dispersion law
(\ref{39}) by substituting $q=\pi - i \lambda^z$ and solving this
equation with respect to $\lambda$ with frequencies $\Omega_k= k
\omega$. As a result, we get

\begin{eqnarray}
\nonumber
\lambda_k^z&=& \ln \left [ \zeta+ \sqrt {\zeta^2-1} \right ]\\
\zeta&=&\frac{\sqrt{D^2+\Omega_k^2}-|D|}{J}-1~,~k=1,2,...  ~~~.
\label{45}
\end{eqnarray}

Since the Fourier components for $S^y_n$ decay in space as $S^z_n$
(except for the static one), we have omitted them. The spatial
decay of the Fourier components of $S^x_n$ can be obtained using
the normalisation condition (\ref{2}), and therefore, for small
deviations from the ground state (\ref{40}), the following
expansion is true: $S_n^x=1-{S_n^y}^2/2-{S_n^z}^2/2+O \left
({S_n^{(y,z)}}^4 \right )$. Substituting here the Fourier
expansion for $S^y_n$ and $S^z_n$, we see that only the product of
terms containing harmonics $k\omega$ and $(m \pm k) \omega$ of
$S^{y,z}$will
contribute to the decay of the $m$th harmonic of $S^x$. We have to
choose the smallest exponent of all possible ones to obtain the
leading order decay rate:

\begin{equation}
\lambda_m^x= \min_{k=0,1,...,m} \left [ \lambda_k^z+\lambda_{m\pm k}^z
\right ]
\label{45a}
\end{equation}

As a result, the following relations have been obtained for the
first five harmonics of the $S^x$ component: $\lambda_0^x=2
\lambda_1^z$, $\lambda_1^x = \lambda_1^z + \lambda_2^z$,
$\lambda_2^x = 2 \lambda_1^z$, $\lambda_3^x= \lambda_1^z +
\lambda_2^z$, $\lambda_4^x = \lambda_1^z + \lambda_3^z$. The
comparison of these theoretical results with the values of
$\lambda$ extracted from the numerical data is given in Table I.
\newpage


\begin{table}
\caption{Numerically and analytically computed values of the decay
exponents $\lambda_k$ of Fourier $k=0,1,..,4$ harmonics for
different spin components. Parameter values are the same as in
Fig. \ref{fig9}.} \label{tab1}
\begin{tabular}{c|c|c|c|c}
Order, $k$& \multicolumn{2}{c|}{$\lambda_k^x$} &
\multicolumn{2}{c}{$\lambda_k^{y,z}$} \\
\cline{2-5} & $~\mbox{numerical}~ $ & $~\mbox{analytical}~ $ &
 $~\mbox{numerical}~ $ & $~\mbox{analytical}~ $ \\
\hline
0     & ~3.5914~ & ~3.5903~ & ~3.1846~ & ~3.1856~ \\
1     & ~4.9463~ & ~4.8055~ & ~1.8067~ & ~1.7951~ \\
2     & ~3.5859~ & ~3.5903~ & ~2.9979~ & ~3.0103~ \\
3     & ~4.8499~ & ~4.8055~ & ~3.5597~ & ~3.5690~ \\
4     & ~5.4393~ & ~5.3641~ & ~3.9195~ & ~3.9309~
\end{tabular}
\end{table}

As can be seen from this Table, the agreement between the
numerical and  analytical values of $\lambda$ decreases with the
order of the Fourier components, which we think is due to the
smallness of the higher order components.

%
%
\section {Two-dimensional lattice with easy-plane anisotropy}

Finally, we briefly consider a two-dimensional system, namely an
easy-plane ferromagnet with  nearest-neighbour exchange
interactions. We have numerically simulated the Landau-Lifshitz
equations for this system:
\begin{eqnarray}
{\dot S}_{mn}^x&=&
\nonumber
J_y S_{mn}^z \left
(S_{m-1,n}^y+S_{m+1,n}^y+S_{m,n-1}^y+S_{m,n+1}^y\right )\\
\nonumber &-& J_z S_n^y
\left ( S_{m-1,n}^z+S_{m+1,n}^z+S_{m,n-1}^z+S_{m,n+1}^z \right) \\
\nonumber &-&2D S_{mn}^y S_{mn}^z \\
{\dot S}_{mn}^y&=&
\nonumber
J_z S_{mn}^x \left (S_{m-1,n}^z+S_{m+1,n}^z+
S_{m,n-1}^z+S_{m,n+1}^z\right )\\
\nonumber
&-&J_x S_{mn}^z \left ( S_{m-1,n}^x+S_{m+1,n}^x+S_{m,n-1}^x+S_{m,n+1}^x
\right) \\
\label{47} &+&2D S_{mn}^x S_{mn}^z \\
{\dot S}_{mn}^z&=&
\nonumber
J_x S_{mn}^y \left (S_{m-1,n}^x+S_{m+1,n}^x
+S_{m,n-1}^x+S_{m,n+1}^x\right  )\\
\nonumber &-&J_y S_{mn}^x
\left ( S_{m-1,n}^y+S_{m+1,n}^y+ S_{m,n-1}^y+S_{m,n+1}^y\right)
\end{eqnarray}
\noindent using again a fourth-order Runge-Kutta scheme with
various initial spin configurations. The results of our
simulations to some extent are similar to the one-dimensional
problem. In Fig. \ref{fig11}, we show the simplest possible
configurations of breathers
%
%
\begin{figure}[htb]
\vspace{2.0pt}
\centerline{\psfig{file=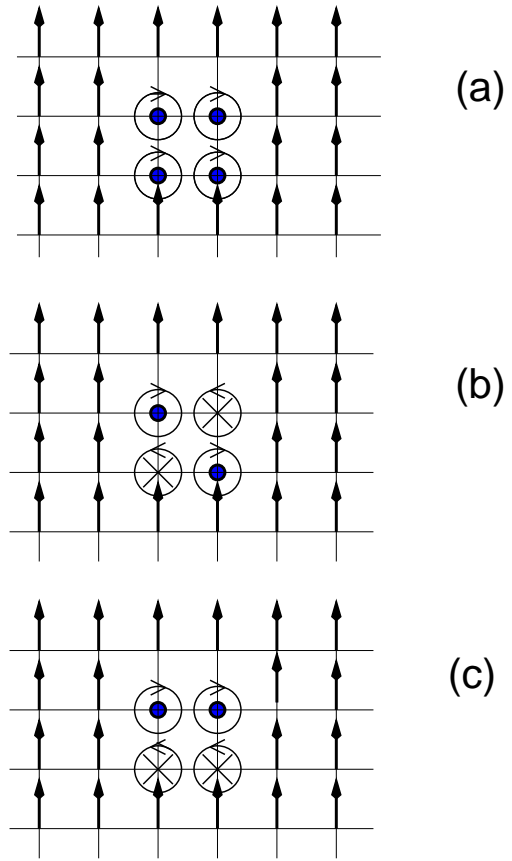,width=4.50in,angle=-90}}
\vspace{5pt}
\caption{Schematical representation of some possible
configurations of discrete breathers with four precessing spins.}
\label{fig12}
\end{figure}
\noindent which involve four `out-of-plane' precessing spins. No
stable breathers with one precessing spin are possible, similarly
to the one-dimensional model. Also, there are no stable breathers
with two or three precessing spins (at least in the limit of small
$J$). Among the three possible stable configurations shown in Fig.
\ref{fig12}, the first one [see panel (a)] corresponds to four
spins precessing parallel and in phase, similar to its
one-dimensional counterpart. The second two configurations do not
have analogues in the one-dimensional case, but have also similar
parity properties. The cases shown in Fig. \ref{fig12}(b,c)
represent breathers with two spins precessing around the $Z$ axis
in the positive direction (marked by dots) and two spins
precessing around the negative direction (marked by crosses).

Simulations have been performed on a lattice of $150 \times 150$
spins with $J=0.11$ and $D=-1$.

%
%
\section {Summary and discussions}

Summarising, we have considered discrete breathers in different
(easy-axis and easy-plane) classical ferromagnetic spin lattices
as essentially discrete objects. We have shown systematically how
to implement the anti-continuum limit for different types of
magnetic lattices (different types of anisotropy). Depending on
the type of anisotropy, discrete breathers have properties similar
to breather solutions for other nonlinear lattices. In the case of
an easy-axis anisotropy the breather solution appears in the gap
of the linear magnon band as do the breathers of the
Klein-Gordon-type models. In the easy-plane case, there is no gap
in the linear band and the breather frequency lies above the band;
these breathers resemble the breathers of the FPU-type chains
(also known as the Sievers-Takeno\cite{st88prl} modes).

The concept of the anti-continuum limit helps us, first, to show
rigorously the existence of discrete breathers, and, second, to
compute the breather solutions numerically. The existence proof
has been performed for the one-dimensional $XYZ$ Heisenberg
ferromagnetic chain with a strong exchange ($J_{x,y} \ll J_z$).
The proof can be easily generalised to the presence of an
easy-axis ion anisotropy and to larger lattice dimensions. The numerical

continuation of the
solutions from the AC limit has been done with the help of a
Newton iteration scheme. Note that so far\cite{ls99pr} only
breathers with one nonzero Fourier component in time (\ref{38a})
have been studied, due to the fact that it is much easier to treat
them both numerically and analytically. The solutions we have
studied allow the infinite number of harmonics, as in $XYZ$ model,
for example.

Why is it important to study discrete systems if the continuum
approximation can give an analytical solution? First of all, it is
known that the breathers are nongeneric for most continuous
models\cite{sk87prl}. Therefore many systems may be incorrectly
referred to as those which do not possess breathers. We
demonstrated this circumstance for the easy axis ferromagnet,
where the slightest exchange anisotropy in the hard plane leads to
a loss of breathers in the continuum model, but not in the case of
a spatial lattice. In addition we have obtained breather solutions
for easy plane ferromagnets which have simply no continuum analog.
This is due to the fact that the spins in the center of the
excitation precess around a tilted axis leading to a local tilt of
the magnetisation.

Finally, we would like to address some important unanswered
questions in this area. The first problem is how to treat quantum
spin lattices (e.g., when the total spin is too small to treat the
lattices classically) and what is the quantum analogue of the spin
breather. Another important question is the breather's mobility.
So far, there is no rigorous existence proof for moving
breathers\cite{fk99pd}, however, Lai and Sievers \cite{ls99pr}
have obtained some numerical results with highly mobile spin
breathers. Since their results are concerned only with breathers
with one Fourier component in time, it is still questionable
whether breathers with an infinite number of harmonics can freely
propagate along the lattice.

\acknowledgements

We thank A. S. Kovalev, B. A. Ivanov, and A. M. Kosevich for useful
discussions.

%
%


\begin{thebibliography}{99}
\bibitem{note1}
Consider a decoupled spin ${\bf S}= (S^x,S^y,S^z)$. In the
action-angle variables $(I,\theta)$, the frequency
of the periodic orbit around the stable groundstate equals
$\omega=dE/dI$,
where $E$ is energy of the spin and the dynamics of the associated angle

is given by $d \theta /dt=\omega(I)$. The the periodic orbit with action

$I_0$
is {\it anharmonic} if $d\omega/dI(I_0) \neq 0$.
\end{thebibliography}
\end{document}